# *On Performance Measures for Infinite Swapping Monte Carlo Methods*

J. D. Doll[1] and Paul Dupuis[2]

[1]Department of Chemistry

Brown University

Providence, RI 02912

[2]Division of Applied Mathematics

Brown University

Providence, RI 02912

October 15, 2014

*Abstract*

We introduce and illustrate a number of performance measures for rare-event sampling methods. These measures are designed to be of use in a variety of expanded ensemble techniques including parallel tempering as well as infinite and partial infinite swapping approaches. Using a variety of selected applications we address questions concerning the variation of sampling performance with respect to key computational ensemble parameters.

***I. Introduction:*** In the present work we introduce and illustrate a number of performance measures for rare-event sampling methods. Of general use in a variety of tempering efforts, we seek to apply these measures to recently introduced infinite (INS) and partial infinite swapping (PINS) approaches.[1-5] As a secondary objective we seek to clarify further the links between those INS/PINS approaches and existing techniques.

Our focus is on performance measures that are generic and broadly representative as opposed to those tied intimately to a particular property. Moreover, we seek measures that are of practical use, simple to implement as well as interpret, and that have a firm theoretical basis. Much of the present discussion is focused on expanded ensemble Monte Carlo methods.[6,7] With the understanding that the generalization to include other (or combinations of) control parameters is both possible and often useful, we restrict the discussion here to situations where the system temperature is the natural variable.

Issues of the type we wish to address are questions related to the variation of sampling performance with the number, range, and distribution of ensemble temperatures. We address such questions through the application of a variety of methods to a number of selected, low-dimensional model problems designed to highlight a specific behavior. Where necessary, more challenging applications are undertaken to confirm the generality of conclusions suggested by these simpler models.

Monte Carlo methods,[6-12] powerful tools in the analysis of many-dimensional systems, are arguably the primary means for the study of the equilibrium properties of complex chemical, biological and materials systems. Their implementation relies on stochastic procedures to sample the configuration space of the associated probability densities and to obtain estimates of equilibrium averages of interest.

Established sampling methods[9-12] are adequate for situations where the underlying equilibrium distribution is well connected. Difficulties arise, however, when the distribution in question involves multiple, effectively isolated regions of importance or "inherent structures".[13] Explicitly, if the relevant probability density is denoted $\mu(x)$, and the average of a property to be computed by $<V>$, then we have

$$<V> = \frac{\int dx\, V(x)\mu(x)}{\int dx\, \mu(x)}.$$

(1.1)

Decomposing the integrations in Eq. (1.1) into terms arising from each of the inherent structures of $\mu(x)$, Eq. (1.1) becomes

$$<V> = \sum_\alpha W_\alpha \langle V \rangle_\alpha,$$

(1.2)

where $W_\alpha$ is the fractional weight of the $\alpha^{th}$ inherent structure

$$W_\alpha = \frac{\int_\alpha dx\, \mu(x)}{\int dx\, \mu(x)}$$

(1.3)

and $<V>_\alpha$ is the average of V over the $\alpha^{th}$ inherent structure,

$$<V>_\alpha = \frac{\int_\alpha dx\, V(x)\mu(x)}{\int_\alpha dx\, \mu(x)}.$$

(1.4)

Although not generally decomposed explicitly in actual applications, we see that computing prototypical equilibrium averages of the type in Eq. (1.2) contains in essence two basic tasks; obtaining estimates of the averages *within* individual inherent structures (Eq. (1.4)) and obtaining estimates of the *relative weights* of those inherent structures (Eq. (1.3)). Averages within a single inherent structure are well suited to established sampling procedures and pose no special difficulties. Estimates of the relative importance of different inherent structures, on the other hand, can prove more challenging.[14,15] When the equilibrium distributions involved become "sparse" and break up into relatively isolated regions, for example, rare-event sampling problems can arise. Unless special care is exercised, movement between these isolated regions within the underlying stochastic procedures can become infrequent and the quality of the associated sampling problematic.

Rare-event sampling problems arise naturally in a variety of applications including, for example, the study of activated processes at low temperatures. Although both the issues involved and the techniques proposed to treat them have greater generality, we focus the present discussion on this particular application. Parallel tempering[16-18] and replica exchange techniques[19] have been found useful in the study of activated processes. These tempering methods utilize an expanded ensemble composed of a product of simple Boltzmann factors at a number of temperatures. Information from the high-temperature portions of the simulation (where the equilibrium distributions are more connected) is used to improve the sampling for the lower temperature portions of the simulation.

Recently, we have introduced a new class of rare-event sampling methods, the INS and PINS techniques.[1-4] In its most complete form, the INS approach can be thought of as the extreme limit of parallel tempering in which all possible permutations of coordinate/temperature associations are attempted at an infinitely rapid rate, a limit that large deviation analysis demonstrates is optimal.[2] This rapid swapping produces an environment in which the natural equilibrium distribution is a more highly connected, symmetrized sum of Boltzmann-like factors as opposed to the single term product form used in conventional parallel tempering.

Sampling and data processing methods required for the implementation of INS and PINS approaches have been developed and presented elsewhere.[1-5] The basic sampling demands involved can be handled using a variety of techniques, including traditional Metropolis,[10] molecular dynamics,[20] and hybrid smart Monte Carlo approaches.[21] The component nature of the densities that arise in INS and PINS approaches adds a variety of new and less familiar features to the problem. We clarify a number of these issues and the methods we have adopted to deal with them in Appendices A-C.

The remainder of the present paper is organized as follows: In Section II we state and examine the basic features of a certain equal occupancy result for tempering simulations. Building on the equal occupancy property, we introduce a convenient performance measure for tempering applications and illustrate its utility using a simple model system. In Section III we extend our initial studies with a series of more challenging applications culminating in general suggestions for ensemble optimization.

***II. Equal Occupancy: The Basics:*** In previous work[3] we have shown that an interesting property of common tempering methods is the uniform or equal occupancy rule. This property is most easily stated in the context of parallel tempering. Following Fig. (1) of Katzgraber, et al.,[22] we label the temperatures in the ensemble by $\{T_n\}$, n=1,$N_T$, and the sequence of temperatures visited by a particular configuration during the simulation by $\{N(m)\}$, m=1,M, where N(m) corresponds to the integer index of the configuration at step m in the calculation. As we have shown[3] the average of N(m) for the simulation, <N>, asymptotically approaches $(N_T + 1)/2$ and the fraction of moves the configuration spends in any one of the various temperature streams (e.g. N(m) = k) is the same, $1/N_T$. Explicitly,

$$<N> = \lim_{M \to \infty} \left( \frac{1}{M} \sum_{m=1}^{M} N(m) \right) \to \frac{(N_T + 1)}{2},$$

(2.1a)

and

$$\lim_{M \to \infty} \left( \frac{1}{M} \sum_{m=1}^{M} \mathbf{1}_{N(m),k} \right) \to \frac{1}{N_T}.$$

(2.1b)

As shown previously,[3] with a slight generalization of the temperature-particle associations the equal occupancy result also holds for infinite and partial infinite swapping methods.

The equal occupancy rule provides a convenient starting point for the discussion of the performance of various implementations of tempering methods.[3] At the most basic level the *absence* of this property signals a qualitative breakdown in the simulation. More generally, the *rate* at which equal occupancy is established is *related* to the core issues involved (such as barrier passage). Unlike measures based on specific properties, however, uniform occupancy has an *a priori* known limiting behavior thus making it a particularly convenient generic probe of sampling performance.

We illustrate the use of uniform occupancy in this Section using a number of simple numerical examples. A convenient model for our purposes is the one-dimensional, double-well potential introduced by Frantz[23] and defined as

$$\frac{V(x)}{\varepsilon} = 3\delta\left(\frac{x}{\sigma}\right)^4 + 4\delta(\alpha-1)\left(\frac{x}{\sigma}\right)^3 - 6\delta\alpha\left(\frac{x}{\sigma}\right)^2 + 1,$$

(2.2)

where

$$\delta = \frac{1}{2\alpha+1}.$$

(2.3)

The parameters $\varepsilon$ and $\sigma$ set the natural energy and length scales, respectively, while the parameter $\alpha$ controls the degree of asymmetry ($\alpha = 1$ symmetric double-well, $\alpha = 0$ single minimum). V(x) has a fixed minimum of zero at $x/\sigma = 1$, a fixed barrier height of $\varepsilon$ at $x/\sigma = 0$, and a variable minimum located at $x/\sigma = -\alpha$. Representative plots of this potential for various values of the asymmetry parameter $\alpha$ are shown in Fig. (1).

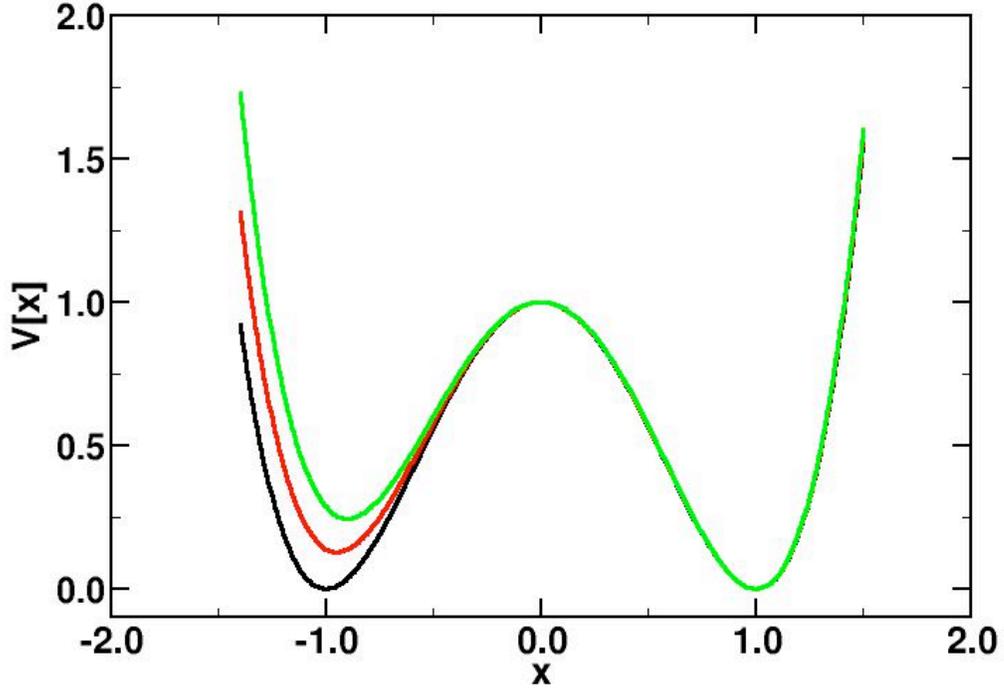

Figure 1: Frantz potential. Shown are plots for α = 1.0 (black), α = 0.95 (red), and α = 0.90 (green). V(x) is in units of ε and x in those of σ.

An essential feature of tempering approaches is their use of expanded equilibrium distributions. In the case of a one-dimensional, two-temperature problem, for example, where the familiar Boltzmann distributions at temperatures $T_1$ and $T_2$ are

$$\pi_1(x) = \exp(-V(x)/k_B T_1),$$

(2.4a)

and

$$\pi_2(y) = \exp(-V(y)/k_B T_2),$$

(2.4b)

the corresponding two-temperature parallel tempering and infinite swapping densities (unnormalized) are

$$\mu_{PT}(x,y) = \pi_1(x)\pi_2(y),$$

and

$$\mu_{INS}(x,y) = \frac{1}{2}\left(\pi_1(x)\pi_2(y) + \pi_1(y)\pi_2(x)\right),$$

(2.5)

(2.6)

respectively. Contours of $\mu_{PT}(x,y)$ and $\mu_{INS}(x,y)$ are shown in Figures (2) and (3) for the Frantz potential for a set of potential parameters chosen with subsequent applications to Ar clusters in mind ($\sigma = 3.405$ Å, $\varepsilon/k_B = 119.8$ K)[24]. Unless otherwise noted, these parameters will be utilized throughout the following discussion. The asymmetry parameter in both Figs. (2) and (3) is $\alpha = 1.0$ and the two representative temperatures involved are $T_1 = 5$ K, $T_2 = 30$ K. To facilitate comparison of the densities, the plot ranges (0.00, 0.03) and contour intervals used in both figures are the same.

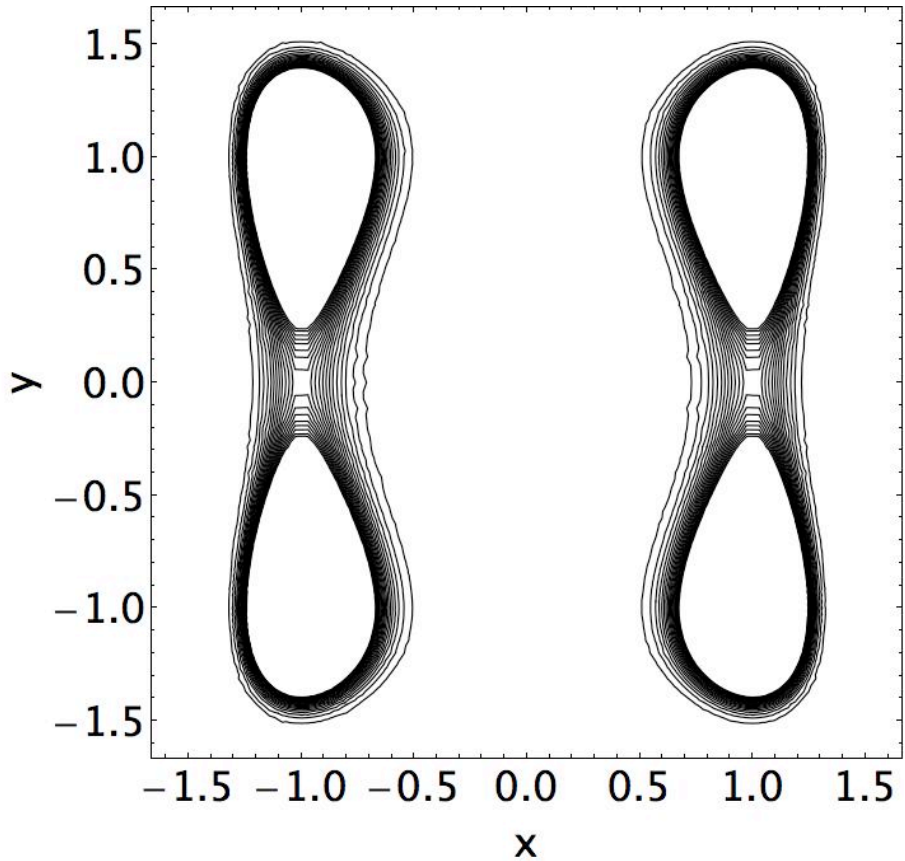

Figure 2: Contour plot of a 2-temperature parallel tempering density ($\alpha = 1.00$, $T_1=10$ K, $T_2=30$ K, $\varepsilon/k_B = 119.8$ K). Coordinate axes are shown in units of $\sigma$.

Comparing Figs. (2) and (3) one sees a significant increase in the connectedness of the infinite swapping versus parallel tempering densities. A more careful analysis reveals that there is, in fact, a slight decrease in density along the y-direction of the infinite swapping result versus its parallel tempering counterpart (the coordinate associated with the higher temperature in Eq. (2.4b)), but a dramatic increase in the density along the x-direction (the coordinate associated with the lower temperature in Eq. (2.4a)). This "Robin Hood" like behavior reflects the tendency of the infinite swapping approach to increase the mobility of the lower temperature portions of the simulation where it is most critically needed at the expense of that of the higher-temperature portions where mobility is less of an issue.

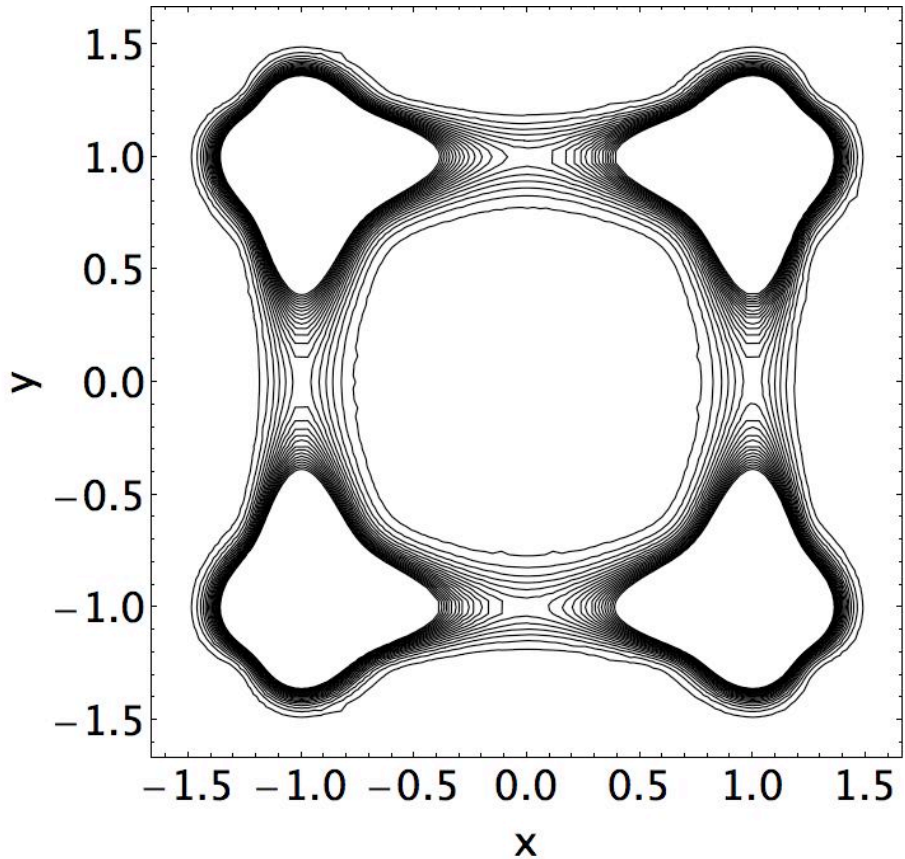

Figure 3: Contour plot of a 2-temperature infinite swapping density ($\alpha = 1.00$, $T_1=10$ K, $T_2=30$ K, $\varepsilon/k_B = 119.8$ K). Plot ranges of Figs. (2) and (3) are the same. Coordinate axes are shown in units of $\sigma$.

Figure (4) shows small portions of sample temperature traces for two-temperature parallel tempering and infinite swapping simulations involving the Frantz potential. Here explicit temperature histories (rather than the integer temperature indices) are plotted. The simulations in Fig. (4) are performed for a slightly asymmetric double well potential ($\alpha = 0.9$) and for slightly different temperatures ($T_1 = 5$ K, $T_2 = 50$ K), but otherwise with the same parameters as those used in Figs. (2) and (3). Unless otherwise stated, the sampling moves in these and all following simulations are performed using smart Monte Carlo methods[21] based on short molecular dynamics segments of 64 steps of 750 a.u. duration. The probability for attempting tempering swaps is taken as 1% for the parallel tempering simulation in Fig (4). Studies for other swap probabilities are discussed later in this Section.

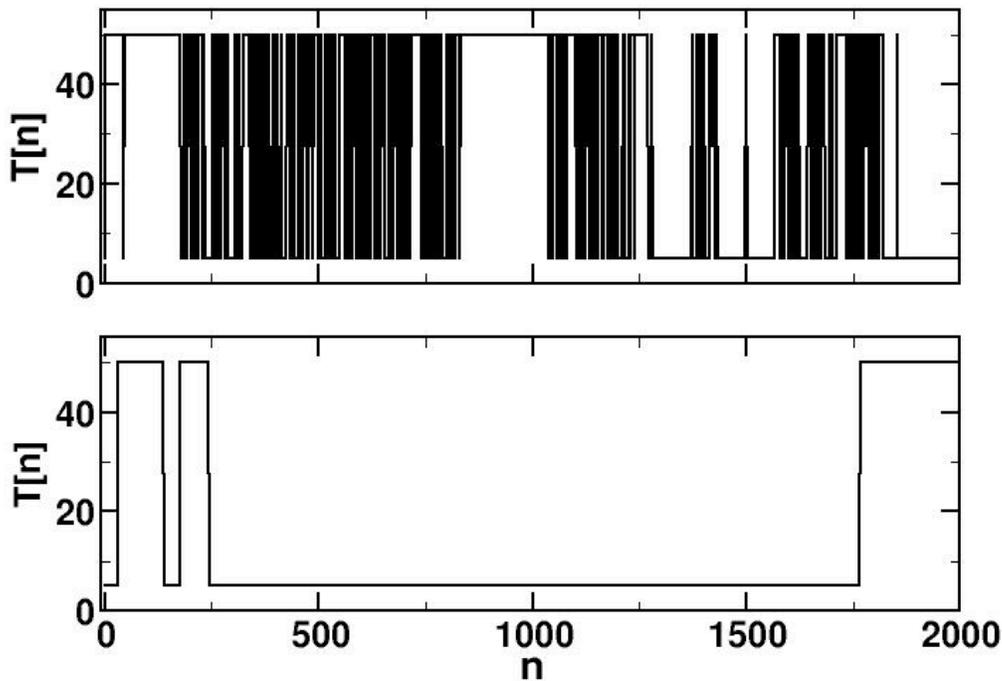

Figure 4: Sample temperature traces for small portions of two-temperature INS (top) and PT (bottom) simulations. Simulations utilized the Frantz potential and the same potential parameters as those in Figs. (2) and (3) ($\alpha = 0.9$, $T_1 = 5$ K and $T_2 = 50$ K.

Appreciable differences in the levels of movement within the tempering ensemble are displayed in the infinite swapping and parallel tempering occupation traces of Fig. (4). These differences reflect the greater rate of information transfer within the computational ensemble for the infinite swapping versus parallel tempering approach.

A convenient device for extracting quantitative information from such simulations is the fluctuation autocorrelation function for the occupation trace, $C_N(s)$, defined as

$$C_N(s) = \frac{<\delta N(m)\delta N(m+s)>}{<\delta N(m)\delta N(m)>},$$

(2.7)

where $\delta N(m)$ is the fluctuation of $N(m)$ about its (known) average value

$$\delta N(m) = N(m) - <N>.$$

(2.8)

For a specified occupation trace, the correlation of the fluctuations separated by s steps, $C_N(s)$, can be estimated as

$$C_N(s) = \frac{\sum_{m\geq 1}\delta N(m)\delta N(m+s)}{\sum_{m\geq 1}\delta N(m)\delta N(m)}.$$

(2.9)

$C_N(s)$ provides a number of convenient measures of the sampling performance. For example, if the fluctuations of $N(m)$ are assumed to be Gaussian in nature, then the sum of $C_N(s)$ over all values of s provides an estimate of the associated correlation length, $N_C$,

$$N_C = \sum_{s=-\infty}^{\infty} C_N(s).$$

(2.10)

In terms of $N_C$ the effective number of independent sample points in an N-point simulation scales as $(N/N_C)$.[25,26] Depending on the problem, $N_C$ is sometimes dominated by the asymptotic decay of $C_N(s)$ thus making the limiting exponential decay rate itself a potentially useful measure of sampling performance. We note that the correlation length is proportional to the asymptotic variance (as defined in, e.g., page 6 of Ref. 27), and that both are inversely proportional to the second derivative of the corresponding large deviation rate function at the mean (here $(N_T+1)/2$).[28] Using the fact that the rate function for INS dominates that of the corresponding parallel tempering scheme, one can argue that INS always reduces correlation length in comparison to any corresponding PT algorithm.

In Figure (5) we plot $C_N(s)$ values extracted from much longer (8 million total smart Monte Carlo points) infinite swapping and conventional parallel tempering simulations for the model system used in Fig. (4). We see $C_N(s)$ results for both sampling methods exhibit an asymptotic exponential decay. We also see that the limiting decay rate is greater for the infinite swapping result than for its parallel tempering counterpart. Further conventional parallel tempering studies of the type in Fig. (5) reveal that the associated asymptotic decay rate first increases and then decreases (but never equals nor exceeds that of INS) as a function of the probability for attempted swaps. The correlation lengths for the various parallel tempering simulations are shown in Fig. (6). Previous studies have found that the relative performance of conventional parallel tempering and infinite swapping methods is more extreme for systems of greater complexity.[1,3]

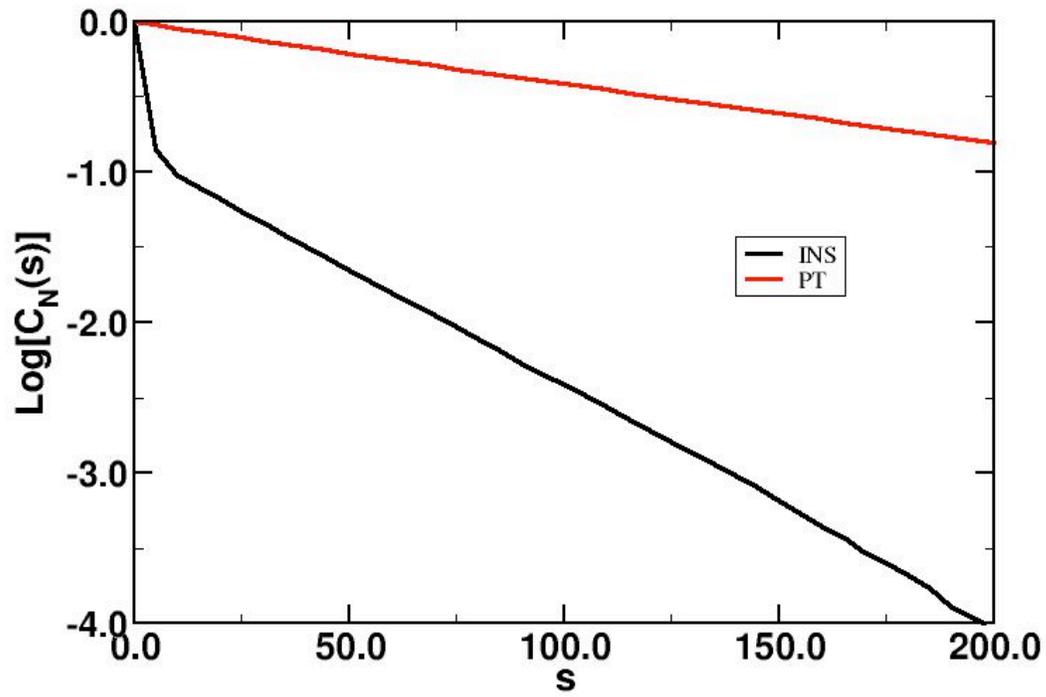

Figure 5: Fluctuation autocorrelation functions $C_N(s)$ for the full INS and PT (1%) simulations of the type depicted in Figure 4.

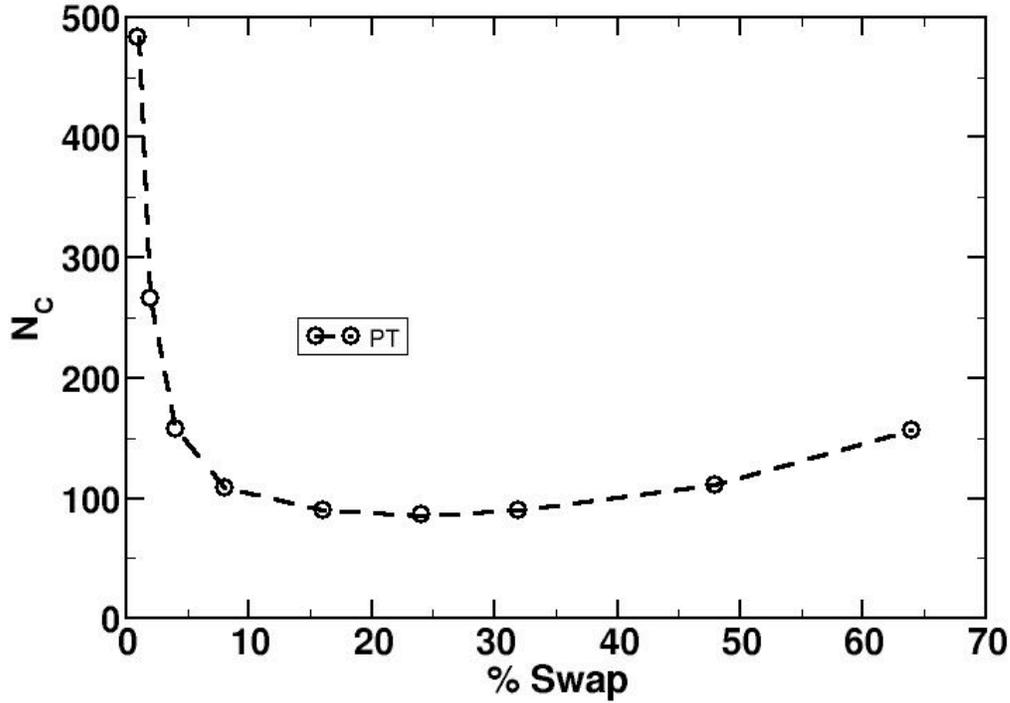

Figure 6: Correlation lengths for PT simulations of the type shown in Fig. (5) as a function of the pair swap attempt percentage. For comparison, the INS $N_C$ value is 53.7

The behavior of the infinite swapping occupation trace in Fig. (4) raises an interesting issue. Specifically, while we see from the figure that the movement within the temperature ensemble has periods of rapid activity, such periods exhibit an "intermittent" or "gated" type of appearance. In Fig. (7) we examine the origin of this telegraph-signal type of behavior by plotting an "indicator" function defined to take on the value of unity if the coordinates of the particles at both temperatures have the same sign and the value zero if they do not. Explicitly, if the two coordinates involved are x and y, the indicator function, **I**(x,y), is defined as

$$\mathbf{I}(x,y) = \begin{cases} 1, & \text{if } xy \geq 0 \\ 0, & \text{otherwise} \end{cases}.$$

(2.11)

One sees that the periods of rapid (slow) tempering swaps are strongly correlated with periods in which the systems at both temperatures are located in the same (different) inherent structure. This means that the rate of approach of the occupancies of the various ensemble temperature streams reflects the key, rare-event barrier crossing events at the heart of the sampling problem. That, plus the fact that such occupancies must approach a known limit, makes uniform occupancy a useful generic probe of sampling performance.

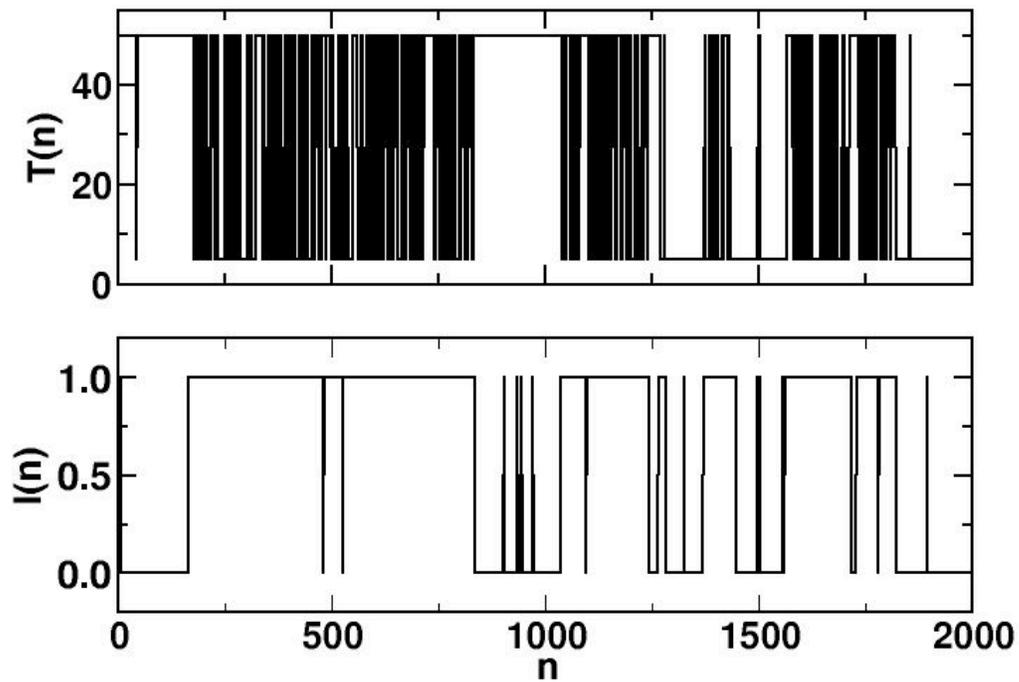

Figure 7: As in Fig (4), but bottom panel here shows indicator (Eq. (2.11)) for both systems to be in same potential well.

***III. Equal Occupancy: General Issues:*** In the present Section we explore a variety of issues related to somewhat more involved aspects of the performance of the infinite and partial infinite swapping methods. Topics of principal concern include the utility of various performance measures and the effects of key computational parameters on subsequent sampling performance. Where appropriate, we utilize examples based on the Frantz potential described in the previous Section for simplicity. Where necessary to probe the generality of such conclusions, we examine related results for more complex systems. Unless otherwise stated all simulations discussed in this Section are obtained using the smart Monte Carlo methods and potential parameters described previously and in Section II.

We begin by addressing a key practical matter, the extent to which the temperature occupation autocorrelation function, $C_N(s)$, serves as a proxy for the calculation of other equilibrium properties. In Fig. (8) we plot fluctuation autocorrelation functions for a number of properties obtained from two-temperature infinite swapping simulations for the Frantz potential[23] with the parameters outlined previously ($\alpha = 0.90$). These simulations utilize $8.4 \times 10^6$ sample points, $T_1=10$ K, and $T_2 = 50$ K. These plots show autocorrelation functions, $C_X(s)$, for the fluctuations in the estimators for the occupation index ($X = N$), for the average potential energy at the lower temperature ($X = V_1$) and for the fractional population of the minor inherent structure ($x < 0$) at the lower temperature ($X = W_{1<}$, c.f. Fig. (7) and Eq. (1.3)). As expected, since all reflect common barrier crossing events, the asymptotic decay rates for the various thermodynamic properties mirror those of $C_N(s)$ making the latter a useful probe of generic sampling performance.

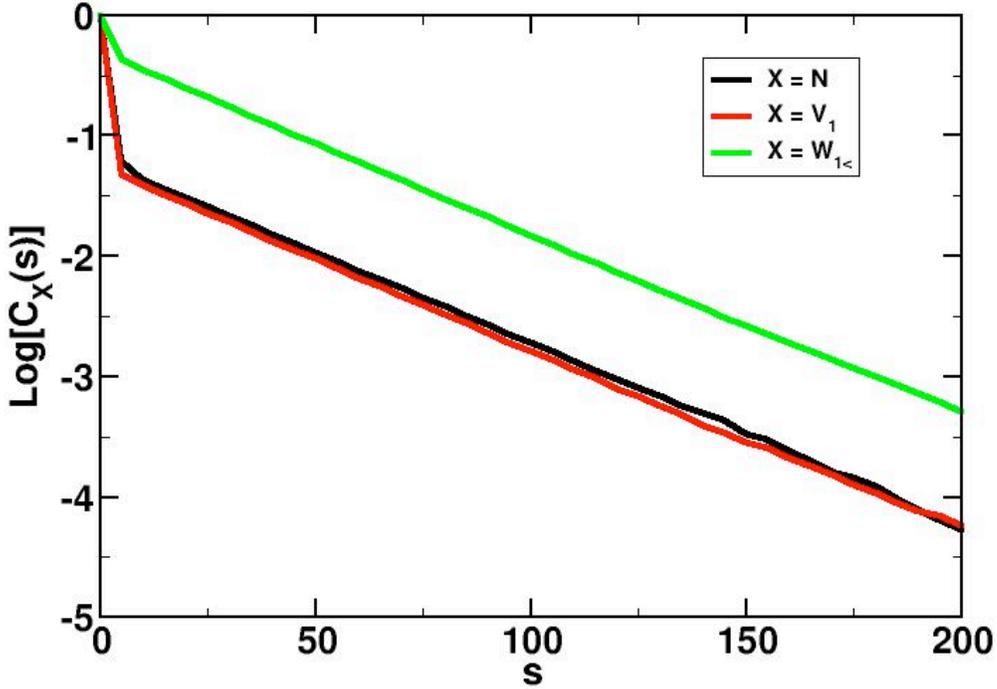

Figure 8: $C_X(s)$ plots for occupation index ($X = N$), potential energy at $T_1$ ($X = V_1$) and fraction of density in the minor inherent structure ($x < 0$) at $T_1$ ($X = W_{1<}$).

A dramatic difference in the dependence on ensemble temperature selection for single versus multiple inherent structure systems is illustrated in Figs. (9) and (10). In these simulations, both based on $8.4 \times 10^6$ sample points, the lower temperature is fixed ($T_1 = 10$ K) while the upper temperature is varied ($T_2 = 20$ K, 30 K, 50 K). Plots of $C_N(s)$ obtained in two-temperature infinite swapping simulations for a single-well Frantz potential ($\alpha = 0$) are shown in Fig. (9). The asymptotic convergence rate of $C_N(s)$ for this barrier-free, single inherent structure system is (a) rapid and (b) does not depend strongly on the maximum temperature in the computational ensemble. In contrast, the decay rates of $C_N(s)$ seen in simulations for the corresponding double-well system ($\alpha = 0.90$) shown in Fig. (10) are (a) slower and (b) display a strong dependence on the maximum temperature used in the computational ensemble. Further analysis reveals that the decay rates in Fig. (10) display a well-defined Arrhenius behavior[29] in which the associated activation energy correlates well with the underlying barrier height for different choices of $\alpha$.

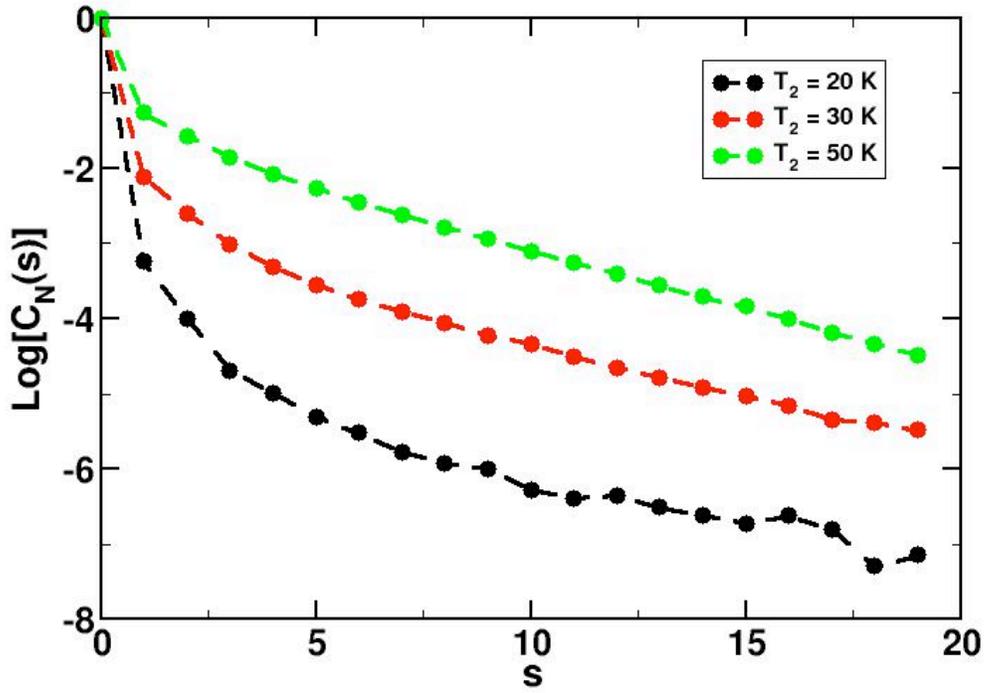

Figure 9: $T_2$ dependence of $C_N(s)$ for 2-temperature Frantz for $\alpha = 0.00$ (single-well), $T_2 = 20, 30, 50$, $T_1 = 10$K.

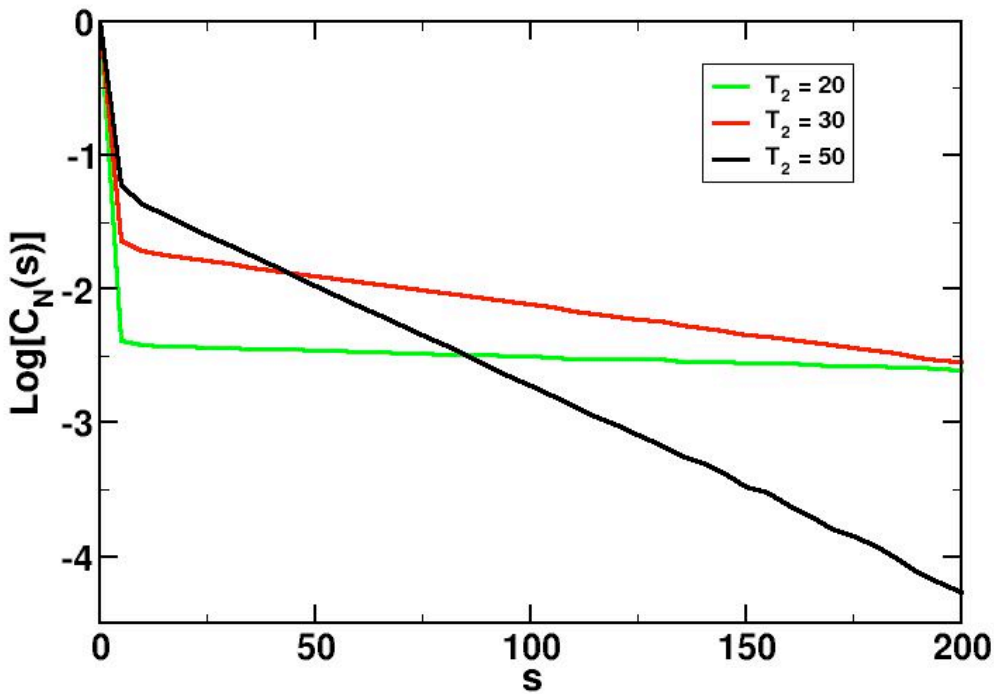

Figure 10: $T_2$ dependence of $C_N(s)$ for 2-temperature Frantz results for $\alpha = 0.90$ (double-well). $N_C$ values for $T_2 = 20$, 30, and 50 K are 194, 87, and 40, respectively. $T_1 = 10K$ for all simulations.

The temperature dependence seen in Fig. (10) carries an important cautionary lesson. As can be seen from those results the reduction of the maximum ensemble temperature in a multiple inherent structure system has two basic effects on $C_N(s)$. The first is to deepen the initial transient drop-off of $C_N(s)$. The second is to reduce the asymptotic decay rate itself, a hallmark behavior of activated, rare-event barrier crossing dynamics. Together these two results mean that if the maximum temperature in the computational ensemble is reduced excessively the onset of the limiting exponential decay of $C_N(s)$ may be missed altogether, or, at minimum, the decay rate in question may become sufficiently slow that obtaining an accurate numerical estimate of it becomes impractical. The computational moral to this story is that when treating multiple inherent structure systems it is important to choose a maximum ensemble temperature that is sufficient to permit both the identification and resolution of the ultimate asymptotic exponential decay of $C_N(s)$.

In Fig. (11) we examine the effects of varying the lower limit of the ensemble's temperature distribution, a companion to the issue considered in Fig. (10). Shown are $C_N(s)$ plots obtained from two-temperature infinite swapping simulations of the Frantz potential ($\alpha = 0.9$) in which the *upper* temperature is fixed ($T_2 = 50$ K) and the *lower* temperature is varied ($T_1 = 5$ K, 10 K, 20 K). We see that in contrast to the effect of varying the maximum temperature, varying the minimum temperature has little effect on the asymptotic decay rate producing instead an overall multiplicative change in $C_N(s)$.

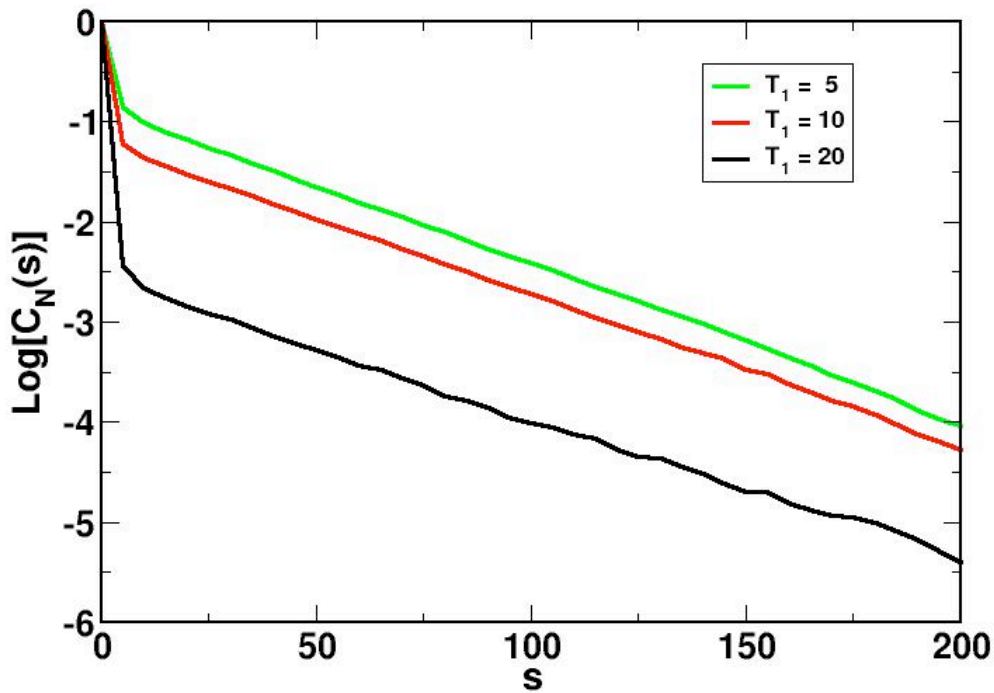

Figure 11: Variation of $C_N(s)$ with respect to $T_1$. $N_C$ values for $T_1 = 5$, 10, and 20 K are 54, 40, and 15, respectively. $T_2 = 50$ K for all simulations. Each $C_N(s)$ curve is obtained using 8.4 x $10^6$ sample points.

We now turn to a series of more involved issues related to sampling performance in multi-temperature applications. Topics of particular interest involve the relative performance of PINS simulations of varying complexity versus that of full infinite swapping and the variation of sampling performance with respect to the number and choice of ensemble temperatures.

Figure (12) shows plots of $C_N(s)$ for simulations of the asymmetric Frantz potential ($\varepsilon/k_B$ = 119.8 K, $\sigma$ = 3.405 Å, $\alpha$ = 0.90) computed with full infinite swapping techniques for varying numbers of ensemble temperatures distributed *uniformly* between a minimum and maximum temperature, 10 K and 30 K, respectively. *Geometrically* distributed temperatures produce qualitatively similar results (not shown). Because differences in the limiting decay rates of the various INS simulations are of interest, the simulations of Fig. (12) utilize a relatively large number of sample points (6.7 x $10^7$), points obtained using smart Monte Carlo methods[21] of the type discussed in Section II. In Table I we list estimates of the limiting slopes of the different simulations (obtained from least squares fits of the results of the type shown in Fig. (12) for the region s > 100) along with estimates from Eq. (2.10) of the associated correlation lengths, $N_C$. The $N_C$ estimates in Table I are based on explicit sums of $C_N(s)$ values over the range of $|s| \leq 100$ and extrapolations of least squares exponential fits for $|s| > 100$.

As seen in both Fig. (12) and Table I, the computed $C_N(s)$ results display a well-defined exponential decay and an interesting dependence on the number of ensemble temperatures. There is a monotonic *worsening* in the asymptotic decay rate but an overall *improvement* in the correlation length with an increasing number of temperatures. The resolution of this apparent contradiction, readily seen in Fig. (12), is that while the asymptotic rate of decay worsens with increasing values of $N_T$, the portion of the correlation length that it contributes decreases. As a practical matter, it should be noted that while the correlation length steadily decreases with larger $N_T$ values, the overall effort involved increases, thus making the general selection of the computationally "optimal" number of ensemble temperatures an application-dependent compromise. We also note that there is a difference in the performance of the geometric and uniform temperature ensembles, a difference we will explore more fully below.

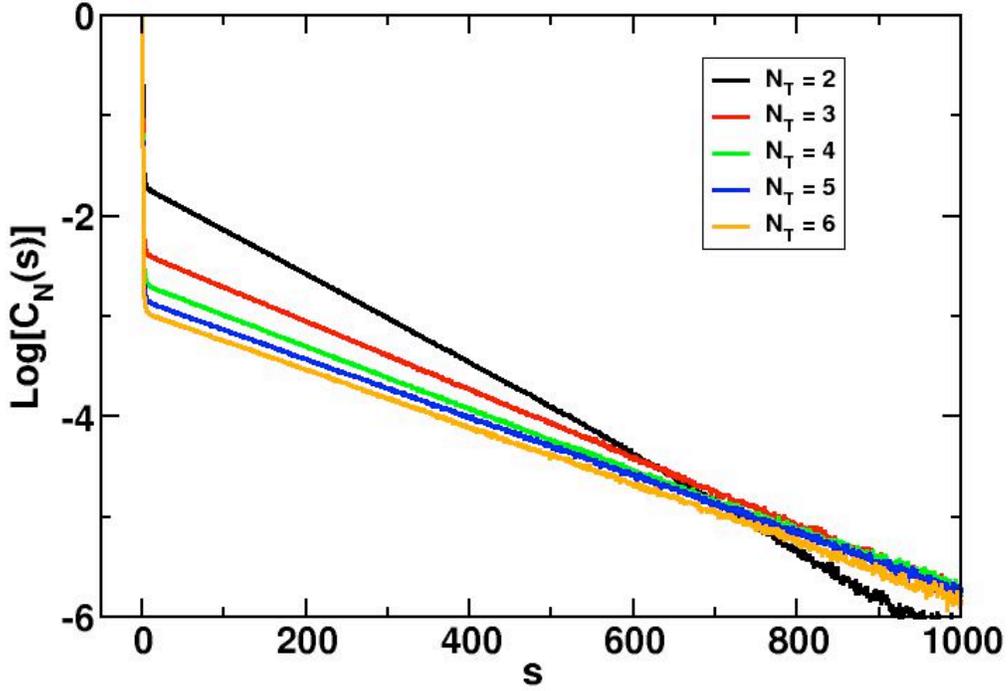

Figure 12: Variation of $C_N(s)$ for INS simulations of an asymmetric double-well potential (a = 0.90) with respect to the number of temperatures, $N_T$, distributed uniformly between 10 K and 30 K. Each $C_N(s)$ curve is obtained using 6.7 x $10^7$ sample points.

================================================================

**Table I:**

Limiting slopes (S) and associated correlation lengths ($N_C$) for the uniform (U) and geometric (G) ensemble INS simulations of the type in Fig. (12) as a function of the number of temperatures ($N_T$) in the simulation.

| $N_T$ | $S(N_T)*1000$ | | $N_C(N_T)$ | |
|---|---|---|---|---|
| | U | G | U | G |
| 2 | 4.60 | 4.60 | 81.2 | 81.2 |
| 3 | 3.38 | 3.31 | 55.8 | 60.0 |
| 4 | 3.02 | 2.85 | 45.1 | 53.0 |
| 5 | 2.90 | 2.71 | 41.2 | 48.4 |
| 6 | 2.87 | 2.55 | 37.2 | 46.5 |

================================================================

It is both possible and useful to develop a simple model for the asymptotic rate of relaxation of spontaneous fluctuations in the populations of various inherent structures of the infinite swapping density. Within a transition state like picture,[30] a component density of the type in Eq (2.6) suggests that the relaxation rate for a double-well system will consist of a sum of Boltzmann-like factors for each of the temperatures in the ensemble. Assuming that the pre-exponentials involved are common and not strongly temperature dependent, one expects the relaxation rate to be of the form of a (constant) pre-exponential factor times the average of the corresponding exponential factors over the set of temperatures in the tempering ensemble. Explicitly, if the temperatures in the ensemble are denoted $\{T_n\}$, $n=1,N_T$ and $E_a$ is the common activation energy, then the effective rate constant for this simplified model of relaxation, $k_{eff}$, should behave as

$$k_{eff} \sim \left(\frac{1}{N_T}\right)\sum_{n=1}^{N_T}\exp(-E_a/k_B T_n).$$

(3.1)

Although the explicit numerical value of $k_{eff}$ will contain a myriad of details involving the sampling process, the hope is that Eq. (3.1) contains enough of the essence of the problem to capture qualitative differences in *relative* decay rates of different ensembles. As a test of this idea we present in Table II the results of this simple, heuristic model applied to the uniform and geometric ensemble results outlined above. Although the quality of the model's fit could likely be improved by treating the activation energy as an adjustable parameter, for simplicity we have chosen to utilize a value equal to the barrier height ($E_a/k_B = 119.8$ K) in the results of Table II. While imperfect, the model appears to capture important aspects of the $N_T$ dependence of the various convergence rates. In particular, it correctly predicts the observed decrease in the asymptotic rate with increasing numbers of ensemble temperatures seen for both ensembles. Surprisingly, given its simplicity, the model also predicts that the asymptotic rates for the geometric ensemble for a given temperature are smaller than those for a uniform distribution for the same number of overall temperatures. The dependence of the asymptotic rate of convergence on the details of the computational ensemble's temperature distribution is an important practical issue and one we will consider in greater detail later in this Section.

===============================================================================

**Table II:**

Ratios of limiting slopes of Table I relative to $N_T = 2$ values for the uniform (U) and geometric (G) ensemble INS simulations versus the corresponding results of the model (M) discussed in the text.

| $N_T$ | $S_U(N_T)/S_U(2)$ | $M_U$ | $S_G(N_T)/S_G(2)$ | $M_G$ |
|---|---|---|---|---|
| 2 | 1 | 1 | 1 | 1 |
| 3 | 0.735 | 0.757 | 0.720 | 0.702 |
| 4 | 0.658 | 0.679 | 0.621 | 0.592 |
| 5 | 0.630 | 0.641 | 0.589 | 0.537 |
| 6 | 0.625 | 0.618 | 0.556 | 0.504 |

===============================================================================

The factorial growth of the computational demands of the complete INS approach with ensemble size places limits on its practical application. As we have shown previously,[1-4] various partial infinite swapping approaches are effective. Using occupation based performance measures we now examine the quality of such PINS approaches and their dependence on a number of simulation parameters. In particular, we examine the performance of multi-temperature simulations in which the number of ensemble temperatures is fixed and the block sizes in the dual-chain sampling method are varied (Figs. (13) and (14)) and in simulations in which the block sizes in the sampling are fixed and the number of temperatures is varied (Figs. (15) and (16)). Figure (13) shows plots of $C_N(s)$ results for the same six-temperature ensemble, same potential, and same smart Monte Carlo procedures used in Fig. (12). In Fig. (13), however, results are obtained using dual-chain PINS simulations with varying block sizes. Partial swapping results are compared with those from full INS methods, something that is feasible for the small number of temperatures involved. Results are shown for PINS-2 simulations (sampling chains with block structures of 1-2-2-1/2-2-2) and PINS-4 simulations (sampling chains with block structures of 2-4/4-2) and for full, six-temperature INS simulations. We see that the INS and PINS results in Fig. (13) are quantitatively quite similar with only relatively minor initial transient decay differences being visible. The results of Fig. (14) explore the same issues as in Fig. (13), but with 18 rather than six temperatures uniformly distributed over the interval from 10-30 K. Although the differences in the various PINS results are somewhat greater than those seen in Fig. (13), we see that the limiting asymptotic decay rates are effectively identical for all simulations. The results of Figures (13) and (14) suggest that PINS approaches can capture substantial levels of the ultimate performance potential of complete infinite swapping techniques in a practical manner. As companions to the studies of Figs. (13) and (14), Figs. (15) and (16) examine the sampling performance of simulations using a fixed PINS block structure and ensembles of varying sizes. Ensembles consist of varying numbers of temperatures (indicated in the figures) uniformly distributed over the 10-30 K interval. Shown are $C_N(s)$ results for the same workhorse asymmetric potential used in studies discussed above and obtained using both PINS-2 (Fig. (15)) and PINS-6 simulations (Fig. (16)).

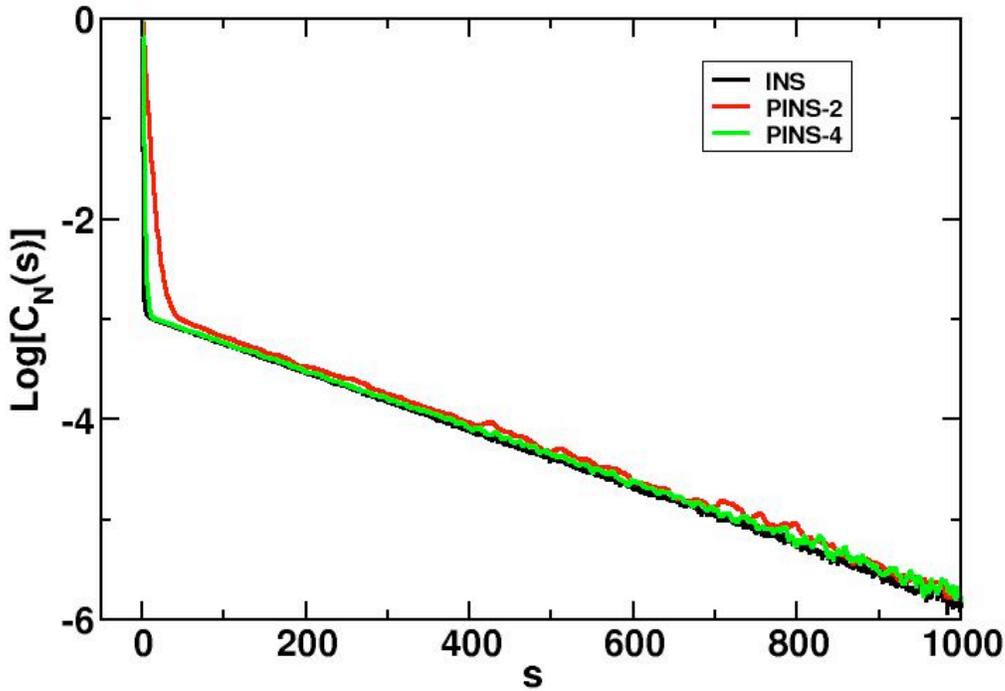

Figure 13: $C_N(s)$ results for various PINS and INS simulations for the same asymmetric system and six temperature ensemble described in Fig. (12) ($\alpha = 0.90$, $T_{min}= 10$ K, $T_{max}= 30$ K, $\varepsilon/k_B = 119.8$ K, and $\sigma = 3.405$ Å). Each $C_N(s)$ simulation is based on $6.7 \times 10^6$ Monte Carlo points. The $N_C$ values for the PINS-2, PINS-4 and INS results are 53, 41, and 37, respectively.

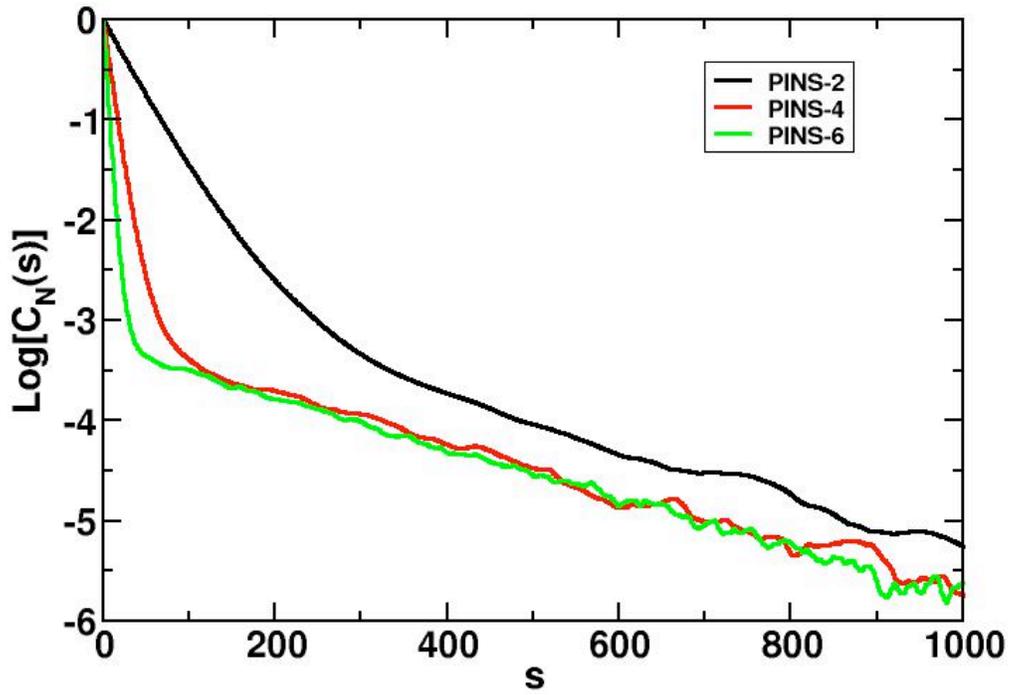

Figure 14: As in Fig. (13), but with 18 ensemble temperatures and PINS simulations based on major block sizes of 2, 4, and 6 temperatures. Each $C_N(s)$ simulation is based on $6.7 \times 10^6$ Monte Carlo points. The $N_C$ values for the PINS-2, PINS-4 and PINS-6 results are 163, 63, and 44, respectively.

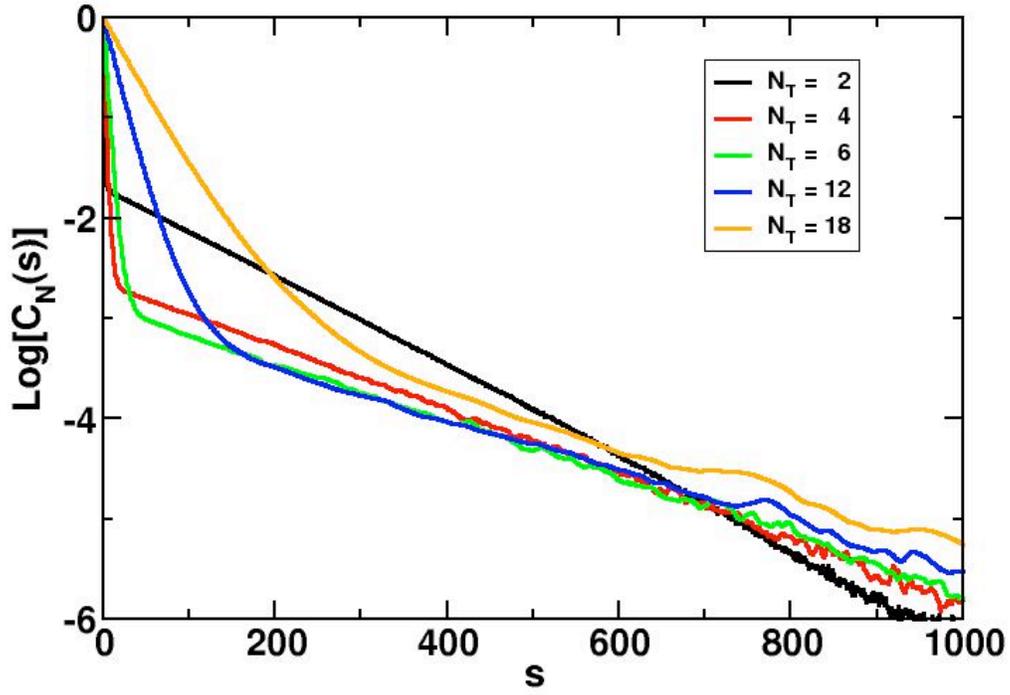

Figure 15: $C_N(s)$ results for PINS-2 simulations for the same system described in Fig. (12) ($\alpha = 0.90$, $T_{min} = 10$ K, $T_{max} = 30$ K, $\varepsilon/k_B = 119.8$ K, and $\sigma = 3.405$ Å). The number of temperatures, $N_T$, ranges from 2-18 as indicated. Each $C_N(s)$ simulation is based on $6.7 \times 10^6$ Monte Carlo points. The $N_C$ values for $N_T =$ (2,4,6,12,18) are (81,51,53,93,164), respectively.

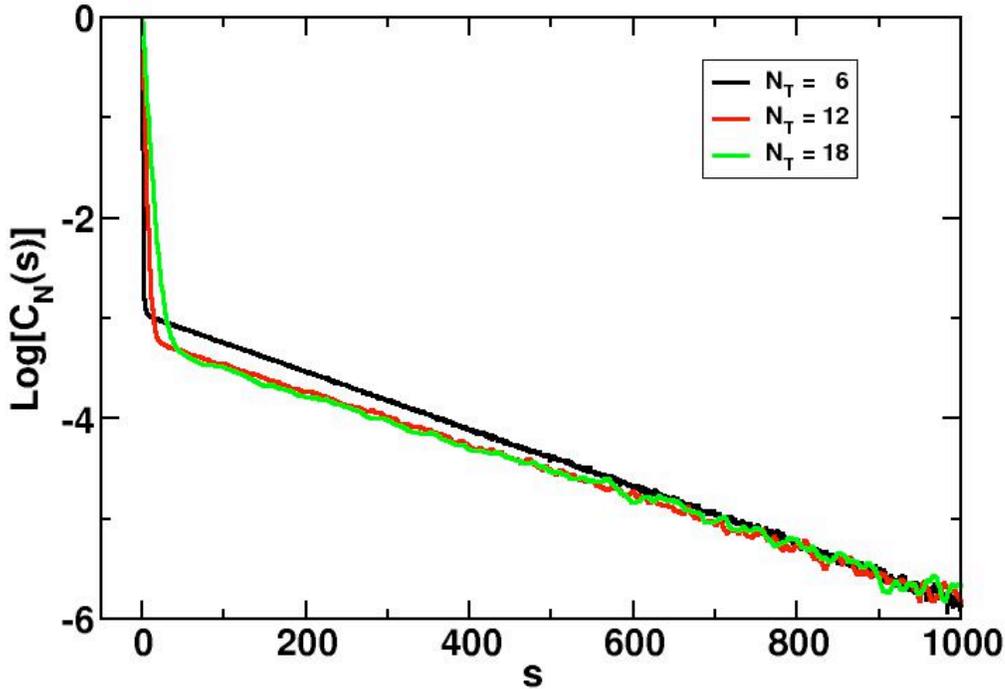

Figure 16: As in Fig. (15), but with PINS-6 simulations. The number of temperatures, $N_T$, ranges from 6-18 as indicated. Each $C_N(s)$ simulation is based on $6.7 \times 10^6$ Monte Carlo points. The $N_C$ values for $N_T$ = (6,12,18) are (37,37,44), respectively.

In the studies described previously, we have been largely concerned with the more qualitative effects of the choice of simulation parameters. We have focused, for example, on the effects of varying numbers of ensemble temperatures as opposed to the details of the distribution of those temperatures. We now turn our attention to this important practical matter, one we have already seen can play a role. The results of Table I, for example, indicate that the manner in which we distribute a fixed number of temperatures over a given interval can affect the sampling performance. A natural question to ask, therefore, is whether or not given the constraints of specified PINS block structures and ensemble size there is an "optimal" choice of temperatures. We address this question with the studies shown in Figs. (17) - (20).

In Fig. (17) we show the $C_N(s)$ values for three, 30-temperature PINS-2 simulations of a Lennard-Jones model of 55-atom argon clusters.[31,32] These simulations utilize the methods and potential parameters described previously. Briefly, conventional Lennard-Jones potential parameters[24] ($\varepsilon/k_B$ = 119.8 K, $\sigma$ = 3.405 Å) and smart Monte Carlo methods[21] based on short (64 steps, 750 a.u./step) molecular dynamics methods are used. A center of mass constraining potential ($R_c = 4\sigma$) and a total of 8 million sample points are used for all simulations. Results are shown in Fig. (17) for three tempering ensembles, including ones in which the 30 temperatures are distributed uniformly, geometrically, and "optimally" across temperatures ranging from 10-50 K. We see from the results of Fig (17) that even for a fixed PINS block and ensemble size, the distribution of ensemble temperatures can have a significant effect on sampling performance. Estimates of the correlation lengths for the $C_N(s)$ results shown in Fig. (17) are 21,000 (optimal), 33,000 (geometric), and 45,000 (uniform).

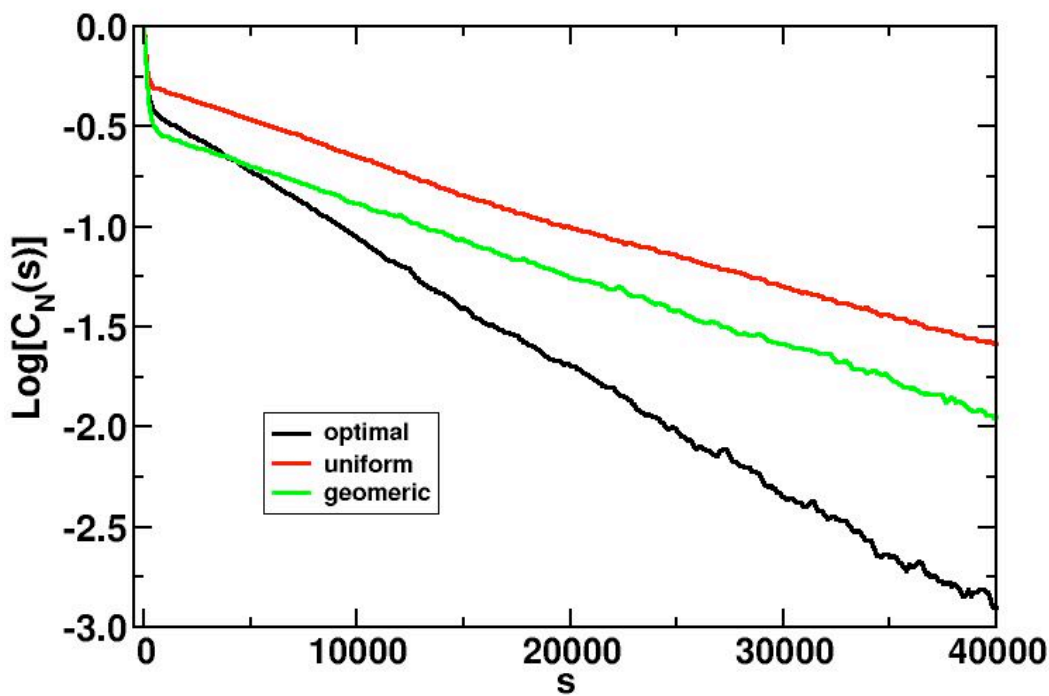

Figure 17: $C_N(s)$ results for 30 temperature PINS-2 simulations of a Lennard-Jones model of $Ar_{55}$. Plots are shown for the uniform, geometric and optimized tempering ensembles discussed in the text.

The various ensembles in Fig. (17) can be conveniently summarized using a parametric plot of the type shown in Fig. (18). Here the black "dots" label the temperatures in the "optimal" ensemble with the abscissa/ordinate of each dot denoting the temperature, $\{T_n\}$, n=1,$N_T$ and index, $\{N(T_n)\}$, N=1,$N_T$, of the corresponding ensemble member. Drawing a horizontal "tie-line" from a given dot to the green or red curves in Fig. (18) produces the corresponding temperature for the geometric or uniform ensembles, respectively. Such a plot facilitates the comparison of ensembles with differing numbers of temperatures.

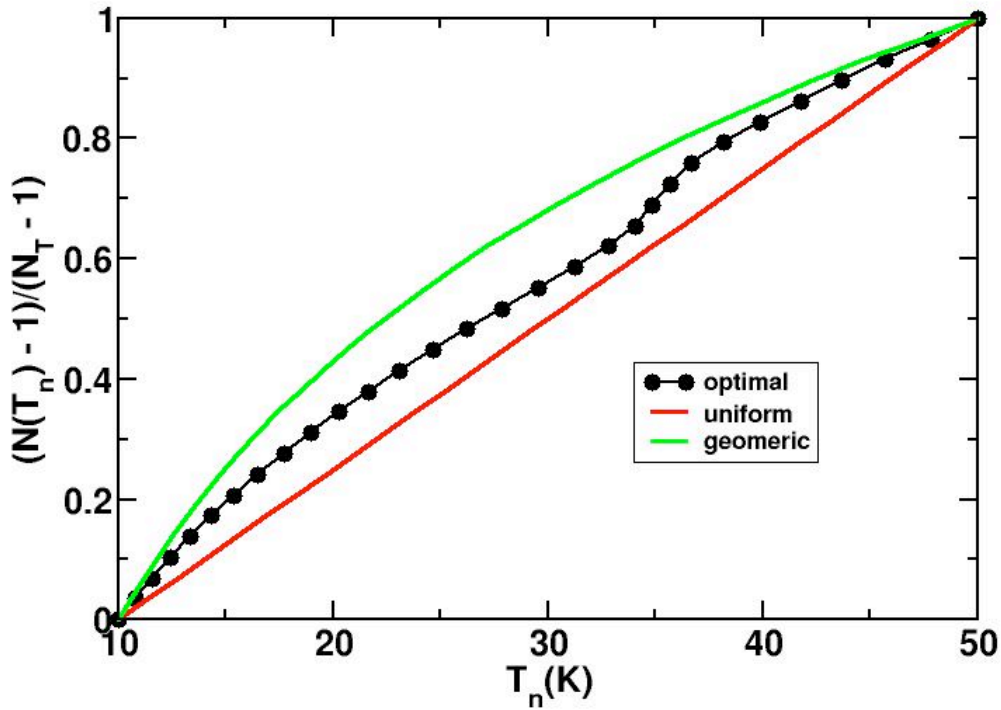

Figure 18: The 30 temperature tempering ensembles used in the Ar$_{55}$ simulations of Fig. (17). The explicit temperatures of the "optimal" ensemble are listed in Appendix D.

The procedure we have used for choosing the "optimal" temperatures in Fig. (18) is suggested in Fig. (19).[3] The underlying probability density in INS and PINS methods is a generalization of Eq. (2.6), i.e. a sum of components corresponding to a fully or partially symmetrized set of coordinate-temperature permutations. The rate of movement within the computational ensemble for the method used in the present work to sample such densities (c.f. Appendix A) varies depending on the relative sizes of the various components in this sum. If the total statistical weight for a given configuration is carried by a single permutation, existing coordinate-temperature associations are largely preserved and little mixing results. If, on the other hand, the statistical weights of the various permutations are more nearly equal, changes in the existing coordinate-temperature associates are likely and mixing will be enhanced. If one uses the relative statistical weights to define a simple information entropy for each of the sampling blocks, heuristically one suspects that choosing ensemble temperatures to render averages of these entropies obtained during the simulation uniform across the tempering ensemble will result in the most efficient transfer of information. Figure (19) shows the block entropies defined in the manner suggested for the three ensembles used in Fig. (17). One sees that for both the uniform and geometric ensembles there is a pronounced "dip" in the block entropies in the 35 K temperature range. The "optimal" ensemble temperatures, established in an iterative manner, tend to be more tightly bunched in this region. This reflects the effects of greater energy fluctuations in this temperature range, a range that contains a pronounced heat capacity maximum for the $Ar_{55}$ system.[33,34] The resulting non-uniform, non-geometric distribution of temperatures produces block entropies that are effectively constant across the ensemble and an asymptotic decay rate greater than that of the other two ensembles. The explicit temperatures used in the optimal ensemble are listed in Appendix D. Those for the uniform and geometric ensembles are readily computed. The present example suggests a general approach for ensemble temperature optimization. Whether the ultimate gain in computational efficiency achieved by such optimization offsets the associated cost will be an application and system-dependent matter. We note in closing that the temperature ensembles generated by the two-temperature PINS block optimization just described work well (i.e. give uniform block entropies) for PINS procedures that use larger block sizes.

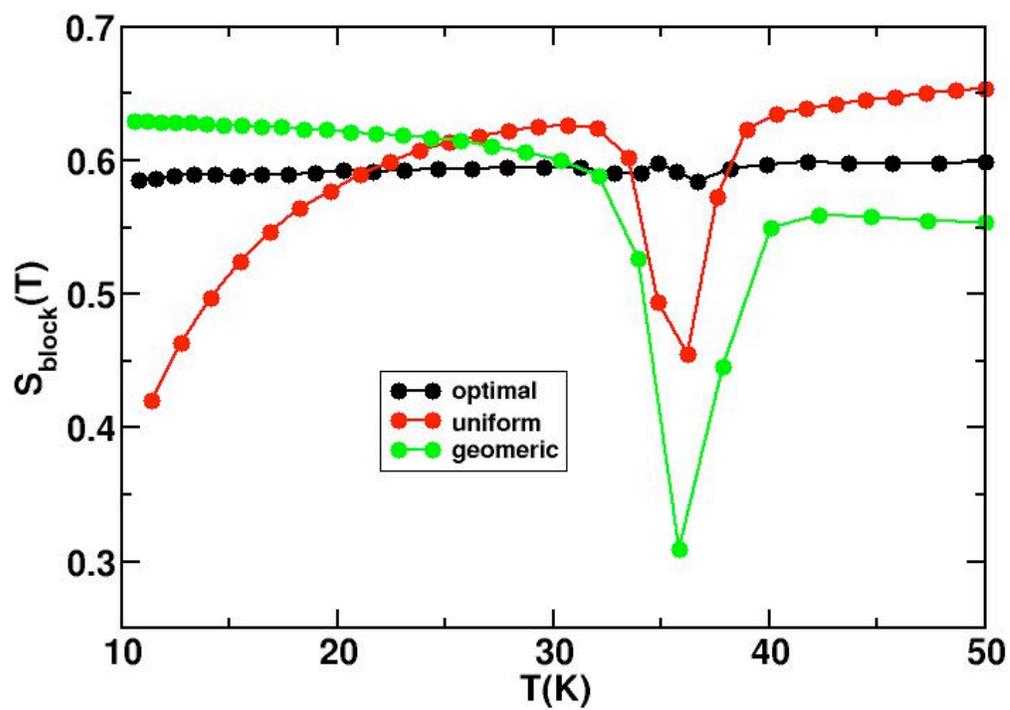

Figure 19: The block entropies for the $Ar_{55}$ simulations of Fig. (17) for the various ensembles.

*IV. Summary:* In the present work we have explored a number of issues related to the performance of a new class of sparse sampling methods, the INS/PINS approaches. A key feature of these infinite swapping methods is their conscious use of symmetry as a tool for dealing with the rare-event problem. Although nominally more complicated than the equilibrium distributions used in typical expanded ensemble applications, the greater connectedness of these symmetrized analogs ultimately justifies their use.

A primary device for our analysis is the equal occupancy property of tempering approaches. As discussed by Katzgraber, et al.,[22] the occupation index, i.e. the history of the movement of a given replica within the temperature ensemble during a tempering simulation, contains valuable information concerning the performance of the associated sampling method. We have shown previously[3] that the fraction of moves a given replica spends in each of the temperature streams within the ensemble is asymptotically equal for parallel tempering as well as for infinite and partial infinite swapping methods. This "equal occupancy" property provides a convenient starting point for the study of the performance of tempering methods. At the most basic level the *absence* of this rigorous property signals a qualitative breakdown in the simulation. More generally, the *rate* at which it is established provides a valuable, general probe of sampling performance.

A primary finding that emerges from the present investigations is the utility of the occupation index autocorrelation function, $C_N(s)$, defined by Eq. (2.9). This function provides a number of convenient performance measures including the fractional occupancy of the various data streams (Eq. (2.1b), the correlation length (Eq. (2.10)), and an asymptotic exponential decay structure that sensitively reflects the problem's underlying rare-event dynamics. Using the occupation autocorrelation function we have examined a number of issues related to the variation of sampling performance with the number, range and distribution of ensemble parameters for a variety of applications in which the natural control parameter is the temperature and have offered general guidelines for the construction of computational ensembles.

*Acknowledgments:* The authors wish to acknowledge support from the National Science Foundation award DMS-1317199 and valuable discussions with Professors David L. Freeman and M. Meuwly and Drs. P. Nyquist, N. Plattner, C. Predescu, and T. Weinreich.

*Appendix A: Sum Sampling Methods:* A common problem in Monte Carlo applications is that of sampling a probability density that is composed of a sum of individual component densities. This issue arises in a somewhat restricted form in the present discussion where the density involved consists of a symmetrized sum of components corresponding to the various temperature-coordinate permutations possible within a specified tempering ensemble. In the present case the component densities are composed of products of Boltzmann distributions and all share a common statistical weights and normalizations. For brevity we denote the total (unnormalized) density, $\mu(x)$, as

$$\mu(x) = \sum_k p_k(x).$$

(A.1)

The details of how densities of the type in Eq. (A.1) arise in the present problem and how they can be processed to extract the properties we seek are described elsewhere.[1-4] The present discussion is focused on the more limited goals of the identification and development of practical methods for their sampling. In situations where the number of component densities is modest (i.e. it is practical to evaluate the total density numerically), several standard sampling approaches are feasible. One is direct Metropolis sampling where coordinate moves are generated using information concerning the ratios of the total density at various locations. Another is the heat-bath method,[35] a two-stage process in which a single component in Eq. (A.1) is first selected in proportion to its density and then that component's density is sampled to produce a new configuration. A typical implementation of this procedure proceeds as follows:
- given a current configuration, $x_n$,
- select a component density, $p_k$, with probability $p_k(x_n)/\mu(x_n)$,
- sample that component density to produce a new configuration, $x_{n+1}$, and
- repeat

The resulting sequence $\{x_n\}$, n=1,N provides a proper sampling of $\mu(x)$. It should be noted that any of a variety of standard procedures can be used for the secondary task of sampling the individual component densities. In general applications, implementation of the heat-bath procedure requires knowledge of ratios of the partition functions for the component densities. However, in the present case all such partition functions are equal and thus their explicit values are unnecessary.

In the present context the factorial growth of the number of terms in Eq. (A.1) means that direct numerical evaluation of the total density, and thus any sampling approach that requires its use, becomes impractical for larger numbers of temperatures. An approach[36] that in principle avoids this difficulty is to sample $\mu(x)$ by repeatedly selecting one of the possible $p_k(x)$ terms *randomly* with the appropriate probability (in this case uniform) and then generating a configuration, $x_n$, from the selected component density. Although only a *single* component density is sampled at a time, it is easy to show that the resulting sequence of configurations, $\{x_n\}$, provides a proper sampling of the *total* density, $\mu(x)$, while avoiding the need for its explicit evaluation. That said, unless the required component sampling can be performed efficiently (e.g. analytically), the single term random selection approach can result in transitions between quite dissimilar component densities, something that can induce extensive warm-up periods in the subsequent configurational sampling. The heat-bath approach, which also samples from a single component density at a time, avoids this difficulty by selecting the new component density in proportion to its relative statistical weight thus assuring that the old and new component densities have some degree of commonality. It does so, however, at the expense of requiring knowledge of the total density, something that is generally unavailable.

What is needed is a sampling approach that combines the desirable commonality feature of the heat-bath method but avoids dealing explicitly with the entire density. A general strategy for proceeding is to modify the basic heat-bath approach by partitioning the entire set of component densities into subsets or "neighborhoods." Rather than selecting from the entire set of permutations, component densities in the modified heat-bath approach are selected from within these neighborhoods, neighborhoods that are themselves sampled as the simulation proceeds. Provided that the neighborhoods involved are suitably defined and generated, the total density can thus be sampled without ever being explicitly evaluated. So formulated, the task of sampling the total density is reduced to the tasks of sampling coordinates from within a given neighborhood and sampling of the neighborhoods for a given configuration.

Because we have control of their size, sampling *within* a given neighborhood presents no special difficulties and can be accomplished using conventional methods (e.g. the heat-bath approach). Sampling *of* the neighborhoods, on the other hand, is a bit more involved. One approach, taken by Ceperley and co-workers in their quantum path integral Monte Carlo treatment of bosonic systems,[35] is to use estimates of the various component densities involved to propose trial neighborhoods and then to use Metropolis acceptance/rejection procedures to assure detailed balance in their selection.

The multi-chain PINS sampling procedure outlined previously[1-4] approaches the task of sampling the density in Eq. (A.1) differently. It retains the component density sampling strategy of the modified heat bath method, but, as illustrated in Appendix C, generates the neighborhoods in such a way that all possible permutation elements are produced in their proper proportion (i.e. with uniform probability). As with single component random sampling, such uniformity eliminates the need for the Metropolis acceptance/rejection step in the neighborhood sampling process. Unlike the single component random sampling approach,[36] however, the multi-chain PINS approach assures commonality between the neighborhoods so generated and does so without making unrefinable assumptions concerning their nature. The utility of the PINS approach for path integral Monte Carlo applications is an open and interesting topic.

*Appendix B: Infinite Swapping and Parallel Tempering:* Using the results of Appendix A it is possible to establish an interesting link between INS and parallel tempering methods. To illustrate this link, let us assume that we are interested in estimating the thermodynamic properties such as the average potential energy of a physical system at a set of temperatures $\{T_n\}$, n=1,$N_T$ using a tempering type approach. We assume further that we denote the average potential energy at the $K^{th}$ temperature as $<V>_K$. To estimate the average imagine performing a thought experiment involving a conventional parallel tempering type simulation, albeit of a somewhat unorthodox design. The simulation is composed of sampling moves of two types; "conventional" moves (each made within a single temperature), and "swaps" that involve attempted tempering pair exchange moves. We begin with an initial configuration for each of the temperatures and then imagine performing a series of attempted parallel tempering swaps between pairs of the temperatures using this *fixed* set of configurations. Rather than a single swap attempt, however, we envision performing a *large* number (N) of such attempts with the *fixed* set of configurations. If N is sufficiently large, then all possible temperature/configuration permutations will be attempted several times. After N such steps, we then make a single, conventional Monte Carlo step in all the coordinates using the configuration/temperature assignments that are those of the final configuration produced by the N swap attempt sequence. This process is then repeated M times and the resulting thermodynamic properties are computed.

If we directly assemble the information produced by the procedure just described, the average potential energy at $T_K$ is given by

$$<V>_K = \frac{\sum_{steps} V(x^{K(step)})}{(MN)},$$

(B.1)

where $x^{K(step)}$ signifies the coordinates for the configuration that is in the $K^{th}$ temperature stream for each of the steps in the simulation. Since the configurations are not changing during the swaps (they are only shuffled around between the different temperatures), it is convenient to express (B.1) somewhat differently. Specifically, we can decompose this expression into a sum over the conventional, single temperature moves and another sum that sweeps over the swaps for that fixed set of configurations. Explicitly,

$$<V>_K = \frac{1}{M}\sum_{m=1}^{M}\left(\frac{1}{N}\sum_{n=1}^{N}V(x^{K(m,n)})\right),$$

(B.2)

where $x^{K(m,n)}$ corresponds to the coordinates for the $K^{th}$ temperature for the $n^{th}$ swap of the $m^{th}$ conventional Monte Carlo step. If N is large (as assumed) then the sum over n includes contributions from all possible permutations multiple times. We can therefore express the inner sum in Eq. (B.2) as the sum over all possible permutations (P) of the potential energy for the $K^{th}$ temperature of that particular permutation of the configurations present after m conventional Monte Carlo moves ($V(x^{K(P,m)})$) times the fraction of times that particular permutation has arisen in the N swap attempts ($N_P/N$). Specifically,

$$<V>_K = \frac{1}{M}\sum_{m=1}^{M}\left(\sum_{P}\frac{N_P}{N}V(x^{K(P,m)})\right).$$

(B.3)

We could implement our hypothetical simulation exactly as described. That is, after each conventional move we could perform a massive set of swaps and use data from those swaps to estimate the inner sum in the above expression. Althernatively, we could perform that average in a *single step* by realizing that the fraction $N_P/N$ that appears will ultimately approach the ratio of the Boltzmann factor for the $P^{th}$ permutation relative to the sum of all the Boltzmann factors for all possible permutations, i.e. the ρ-weights in Refs (1-4). We can thus perform the "infinite" sum over all possible tempering swaps in a single step by suitably processing the information that emerges from the conventional Monte Carlo steps.

After performing the (infinite) sum of the swap attempts, it remains to perform the remaining part of our procedure, a conventional move for the coordinates in each of the temperature streams. If we follow the procedure described above precisely, we would do so by tracking the sequence of permutations, finding the last configuration visited during the swap attempts, and using those configuration/temperature assignments as the starting point for the next conventional MC move. Rather than doing that, however, it is easier to realize that if N is large, the probability of having stopped on any of the various permutations would be in proportion to its corresponding ρ-weight. We can therefore proceed in a statistically equivalent manner by selecting one of the possible permutations of temperature/coordinate associations at random according to its ρ-weight and using the assignments in that randomly chosen permutation to perform the required conventional, single-temperature step. As discussed in Appendix A, such a procedure provides a sampling of the equilibrium distribution that is the *sum* of the component distributions involved, here the symmetrized sum over the possible temperature/coordinate permutations. The combined procedure just described, the ρ-weighted accumulation of data combined with a conventional, single temperature Monte Carlo move of the configurations based on the temperature/coordinate assignments of a randomly chosen permutation, is the implementation of the INS approach outlined in Refs (1-4).

When the number of temperatures becomes large, the number of permutations grows to the point where the full infinite swapping approach becomes impractical. The partial infinite swapping (PINS) approach is designed to retain many of the infinite swapping (INS) ideas, but in a numerically tractable manner. At its core the INS approach works by enforcing the relative Boltzmann weights of the various permutations. This can be done in a single step so long as we can explicitly consider the relative Boltzmann weights of all possible permutations. In the partial swapping approach, the necessary enforcement is done in a multi-step fashion. Rather than considering all possible permutations of the temperature/coordinate associations, in the PINS method the temperature ensemble is partitioned into blocks that each contains distinct subsets of the temperatures in the ensemble. The relative weighting of the permutations *within* each of the blocks is enforced by an INS-like procedure. While this assures that the relative weights of the permutations within each of the PINS blocks is correct, it does not assure that the relative weights of those *between* different blocks is valid. The PINS approach solves this problem by utilizing multiple, distinct PINS blocking arrangements (i.e. multiple "chains") and passing information back and forth between them. As discussed in Appendix C while each of the chains individually produces multiple equilibrium distributions, if the partitionings of the chains and the passage of information between the blocks are suitably designed the only equilibrium distribution the chains have in common is the one that would have been produced by the full infinite swapping approach had we been able to implement it.

*Appendix C: INS, PINS, and Parallel Tempering:* A key task in the infinite swapping approach is the development of a means for sampling the symmetrized combination of all possible temperature-coordinate component densities arising in an expanded tempering ensemble. Methods for accomplishing this task have been presented and documented previously.[1-4] Here we describe the essence of these methods from a somewhat different point of view. The resulting analysis provides a convenient vehicle for discussing INS, PINS and parallel tempering methods and the rate of flow of information within computational ensembles of various designs. For simplicity, the present discussion will consider explicitly the case of a three-temperature ensemble. Generalizations to arbitrary numbers of temperatures will be indicated.

It is convenient to represent the six possible three-temperature products of equilibrium densities as "cells", $\{C_n\}$, n=1,6 defined as

$$C_1 = \begin{pmatrix} \pi_1(x_1) \\ \pi_2(x_2) \\ \pi_3(x_3) \end{pmatrix}, C_2 = \begin{pmatrix} \pi_1(x_1) \\ \pi_2(x_3) \\ \pi_3(x_2) \end{pmatrix}, C_3 = \begin{pmatrix} \pi_1(x_2) \\ \pi_2(x_1) \\ \pi_3(x_3) \end{pmatrix}$$

$$C_4 = \begin{pmatrix} \pi_1(x_2) \\ \pi_2(x_3) \\ \pi_3(x_1) \end{pmatrix}, C_5 = \begin{pmatrix} \pi_1(x_3) \\ \pi_2(x_1) \\ \pi_3(x_2) \end{pmatrix}, C_6 = \begin{pmatrix} \pi_1(x_3) \\ \pi_2(x_2) \\ \pi_3(x_1) \end{pmatrix}$$

(C.1)

In Eq. (C.1) $\pi_i(x_j)$ denotes the equilibrium Boltzmann density for system-j at temperature $T_i$. $C_1$ thus represents a component density $\pi_1(x_1)\pi_2(x_2)\pi_3(x_3)$ in which system-1 is at a temperature $T_1$, system-2 is at a temperature of $T_2$, and so on. One can view the various component densities in Eq. (C.1) as arising from the actions of suitably defined permutation operators. For example, if we define the permutation matrices, $\mathbf{P}_n$, as

$$P_1 = \begin{pmatrix} 1 & 0 & 0 \\ 0 & 1 & 0 \\ 0 & 0 & 1 \end{pmatrix}, P_2 = \begin{pmatrix} 1 & 0 & 0 \\ 0 & 0 & 1 \\ 0 & 1 & 0 \end{pmatrix},$$

$$P_3 = \begin{pmatrix} 0 & 1 & 0 \\ 1 & 0 & 0 \\ 0 & 0 & 1 \end{pmatrix}, \quad P_4 = \begin{pmatrix} 0 & 1 & 0 \\ 0 & 0 & 1 \\ 1 & 0 & 0 \end{pmatrix},$$

$$P_5 = \begin{pmatrix} 0 & 0 & 1 \\ 1 & 0 & 0 \\ 0 & 1 & 0 \end{pmatrix}, \quad P_6 = \begin{pmatrix} 0 & 0 & 1 \\ 0 & 1 & 0 \\ 1 & 0 & 0 \end{pmatrix},$$

(C.2)

and assume these matrices act on the various system's coordinates, then the results of Eq. (C.1) can be summarized as

$$\mathbf{C}_n = \mathbf{P}_n \mathbf{C}_1$$

(C.3)

Using the definitions in (C.1)-(C.3), the rules for combining successive permutation operations can be easily evaluated ($\mathbf{P}_2\mathbf{P}_2 = \mathbf{P}_1$, $\mathbf{P}_2\mathbf{P}_3 = \mathbf{P}_4$, and so on).

Using the definitions above, we can construct probability densities in which the temperature-coordinate associations are fully or partially symmetrized. For example, the fully symmetrized density, $\mathbf{C}_{INS}$, for the three-temperature case is the equally weighted sum of all component densities,

$$\mathbf{C}_{INS} = \left(\frac{1}{6}\right) \sum_{n=1}^{6} \mathbf{C}_n .$$

(C.4)

Using (C.3), we see that $\mathbf{C}_{INS}$ can be written as

$$\mathbf{C}_{INS} = \mathbf{P}_{INS} \mathbf{C}_1 .$$

(C.5)

where the operator $\mathbf{P}_{INS}$ is defined as

$$\mathbf{P}_{INS} = \left(\frac{1}{6}\right) \sum_{n=1}^{6} \mathbf{P}_n .$$

(C.6)

We note that $\mathbf{P}_{INS}$ has the property that it completely symmetrizes any of the possible $\mathbf{C}_n$ densities defined in Eq. (C.1). We also note that $\mathbf{P}_{INS}$ of Eq. (C.6) is given in matrix form (c.f. Eqs. (C.2)) as

$$\mathbf{P}_{INS} = \begin{pmatrix} \frac{1}{3} & \frac{1}{3} & \frac{1}{3} \\ \frac{1}{3} & \frac{1}{3} & \frac{1}{3} \\ \frac{1}{3} & \frac{1}{3} & \frac{1}{3} \end{pmatrix},$$

(C.7)

has the property that $\mathbf{P}_{INS}^2 = \mathbf{P}_{INS}$, and has eigenvalues of $(1,0,0)$. Viewed as a transition matrix, the single unit eigenvalue of $\mathbf{P}_{INS}$ implies that its repeated application will produce a unique limiting distribution while the unit gap between its largest and smallest eigenvalue assures that this limiting distribution is produced in a single iteration.

It is convenient to define two other operators, $\mathbf{P}_A$ and $\mathbf{P}_B$, as:

$$\mathbf{P}_A = \left(\frac{1}{2}\right)(\mathbf{P}_1 + \mathbf{P}_3) = \begin{pmatrix} \frac{1}{2} & \frac{1}{2} & 0 \\ \frac{1}{2} & \frac{1}{2} & 0 \\ 0 & 0 & 1 \end{pmatrix},$$

(C.8a)

$$\mathbf{P}_B = \left(\frac{1}{2}\right)(\mathbf{P}_1 + \mathbf{P}_2) = \begin{pmatrix} 1 & 0 & 0 \\ 0 & \frac{1}{2} & \frac{1}{2} \\ 0 & \frac{1}{2} & \frac{1}{2} \end{pmatrix}.$$

(C.8b)

Unlike their fully symmetrized counterpart, operators $\mathbf{P}_A$ and $\mathbf{P}_B$ only "partially" symmetrize a single product density. For example, the action of $\mathbf{P}_A$ on $\mathbf{C}_1$ is given by

$$\mathbf{P}_A \mathbf{C}_1 = \left(\frac{1}{2}\right)(\mathbf{C}_1 + \mathbf{C}_3),$$

(C.9a)

where $(\mathbf{C}_1 + \mathbf{C}_3)/2$ is the partially symmetrized density $(\pi_1(x_1)\pi_2(x_2) + \pi_1(x_2)\pi_2(x_1))\pi_3(x_3)/2$. The corresponding action of $\mathbf{P}_B$ is

$$\mathbf{P}_B \mathbf{C}_1 = \left(\frac{1}{2}\right)(\mathbf{C}_1 + \mathbf{C}_2).$$

(C.9b)

Using Eqs. (C.7) and (C.8) it is easy to show that

$$\lim_{m \to \infty} (\mathbf{P}_A \mathbf{P}_B)^m = \mathbf{P}_{INS}.$$

(C.10)

In the language of Appendix A, the significance of Eq. (C.10) is that it provides a means for sampling the fully symmetrized density without explicitly dealing with the entire density (i.e. all 3! terms) at once. From a transition matrix point of view, although neither of the partially symmetrized operators $\mathbf{P}_A$ nor $\mathbf{P}_B$ would do so *individually*, the repeated action of the $\mathbf{P}_A\mathbf{P}_B$ product will produce the same, fully symmetrized density as that produced by $\mathbf{P}_{INS}$. This result is again reflected in the eigenvalues of the operators involved with those of $\mathbf{P}_A$ and $\mathbf{P}_B$ each being (1,1,0) while those of the $\mathbf{P}_A\mathbf{P}_B$ product are (1,1/4,0). The multiple unit eigenvalues of $\mathbf{P}_A$ and $\mathbf{P}_B$ imply that neither operator has a unique limiting distribution, whereas the single unit eigenvalue of the $\mathbf{P}_A\mathbf{P}_B$ product indicates that it does. The non-zero second eigenvalue of the $\mathbf{P}_A\mathbf{P}_B$ product means that its repeated application will produce the same limiting distribution as that of the fully symmetrized operator, $\mathbf{P}_{INS}$, while the smaller gap between its unit and second eigenvalue means that it will do so at a slower rate that $\mathbf{P}_{INS}$. This rate reduction is the price paid in the dual chain approach for avoiding the evaluation of the total density. The explicit example just described is the essence of the 1-2/2-1 dual chain PINS sampling approach for three-temperature simulations described in Ref. (1).

From the point of view of Appendix A, Eq. (C.10) means that the dual chain process provides a means for generating all possible temperature-coordinate permutations in their proper relative amounts (i.e. equal). This feature plus the locality associated with the refinable block structure of the dual chain approach provides a convenient means for producing the neighborhoods required for implementation of the PINS sampling procedure.

It is straightforward to generalize the above arguments to N-temperature INS and PINS procedures. In an N-temperature INS process, for example, the relevant extension of Eq. (C.7) is an NxN matrix in which all elements are equal to $1/N$. In the dual-chain PINS approach, the relevant transition matrices for the various PINS chains correspond to block-diagonal forms in which the elements in each of those blocks are equal. Specifically, if the blocks involve M temperatures, the block-diagonal pieces are M x M in dimension and all elements within that block are equal to $1/M$. The gap between the first and second largest eigenvalues of the product of the PINS transition matrices controls the rate of approach to the limiting distribution of the associated dual chain PINS procedure and can be used as a basis for discussing the relative performance of the various PINS approaches.

We close by noting that conventional parallel tempering can be analyzed using the methods outlined above. If there are a total of N ensemble temperatures and at each step in the simulation one attempts a randomly chosen tempering swap of nearest neighbor data streams, then the number of simulation moves required for a given particle to transit an N-temperature ensemble, a number that one expects to mirror the asymptotic relaxation time, scales as $N^3$ for parallel tempering. This result can be understood qualitatively by noting that since the movement of particles between temperature streams is diffusive, the number of swaps required for transit of an N-temperature ensemble will scale as $N^2$. Since the temperature streams to be swapped in parallel tempering are chosen randomly, however, the number of steps required to assure a swap of a *particular* particle is proportional to N thus producing the overall $N^3$ scaling for the transit time of a given particle. In contrast, since the populations of all temperature streams are moved at each step, the transit times for an N-temperature PINS simulation should scale as $N^2$.

*Appendix D: Temperatures of Optimal Ensemble:*

For completeness, we have listed below are the explicit temperatures used in the 30-temperature "optimal" ensemble simulations shown in Fig. (18). The corresponding geometric and uniform ensemble temperatures can be easily generated by direct algebraic means.

==================================================================

Table III

Shown are the 30 temperatures for the optimal ensemble discussed in Section III.

| n | $T_n$ | n | $T_n$ | n | $T_n$ |
|---|---|---|---|---|---|
| 1 | 10.00 | 11 | 20.26 | 21 | 34.88 |
| 2 | 10.77 | 12 | 21.65 | 22 | 35.68 |
| 3 | 11.58 | 13 | 23.11 | 23 | 36.71 |
| 4 | 12.45 | 14 | 24.63 | 24 | 38.15 |
| 5 | 13.37 | 15 | 26.22 | 25 | 39.90 |
| 6 | 14.35 | 16 | 27.87 | 26 | 41.75 |
| 7 | 15.40 | 17 | 29.54 | 27 | 43.70 |
| 8 | 16.51 | 18 | 31.23 | 30 | 45.73 |
| 9 | 17.69 | 19 | 32.85 | 29 | 47.83 |
| 10 | 18.95 | 20 | 34.05 | 30 | 50.00 |